\documentclass[journal]{IEEEtran}
\ifCLASSINFOpdf
  \usepackage[pdftex]{graphicx}
  \graphicspath{{../} {../pdf/} {../jpeg/}}
  \DeclareGraphicsExtensions{.pdf,.jpeg}
\else
  \usepackage[dvips]{graphicx}
  \graphicspath{{../} {../eps/}}
  \DeclareGraphicsExtensions{.eps}
\fi

\usepackage{amsmath}
\usepackage{amssymb}
\usepackage{algorithm}
\usepackage{algpseudocode}

\usepackage{array}
\usepackage{multirow}
\usepackage{booktabs}
\usepackage{lipsum}
\usepackage{bm}
\usepackage{color}

\usepackage{footnote}
\usepackage{tabularx}
\usepackage[flushleft]{threeparttable}
\usepackage[dvipsnames]{xcolor}
\usepackage[normalem]{ulem}

\usepackage{url}

\begin{document}

\bstctlcite{IEEEbib:BSTcontrol}

\title{A Non-Intrusive Low-Rank Approximation Method for Assessing the Probabilistic Available Transfer Capability}

\author{Hao~Sheng,~\IEEEmembership{Member,~IEEE,}
    and~Xiaozhe~Wang,~\IEEEmembership{Member,~IEEE,}
\thanks{This work is supported by Natural Sciences and Engineering Research Council (NSERC) under Discovery Grant NSERC RGPIN-2016-04570.}
\thanks{The authors are with the Department of Electrical and Computer Engineering, McGill University, Montr\'{e}al, QC H3A 0G4, Canada. (e-mail: shenghao@tju.edu.cn, xiaozhe.wang2@mcgill.ca)}
}

\markboth{IEEE Transactions on Smart Grid, ~Vol.~, No.~, ~2018}%
{Sheng \MakeLowercase{\textit{et al.}}:
A Non-Intrusive Low-Rank Approximation Method for Assessing the Probabilistic Available Transfer Capability
}

\maketitle



\begin{abstract}
In this paper, a mathematical formulation of the probabilistic available transfer capability (PATC) problem is proposed to incorporate uncertainties from the large-scale renewable energy generation (e.g., wind farms and solar PV power plants). Moreover, a novel non-intrusive low-rank approximation (LRA) is developed to assess PATC, which can accurately and efficiently estimate the probabilistic characteristics (e.g., mean, variance, probability density function (PDF)) of the PATC. 
Numerical studies on the IEEE 24-bus reliability test system (RTS) and IEEE 118-bus system show that the proposed method can achieve accurate estimations for the probabilistic characteristics of the PATC with much less computational effort compared to the Latin hypercube sampling (LHS)-based Monte Carlo simulations (MCS).
The proposed LRA-PATC method offers an efficient and effective way to determine the available transfer capability so as to fully utilize the transmission assets while maintaining the security of the grid. 
\end{abstract}


\IEEEpeerreviewmaketitle

%
\section{Introduction}

\IEEEPARstart{A}{vailable} transfer capability (ATC) is a crucial index designed to describe how much more electric power (MW) is available to buy or sell in a specified period in a competitive electric power market. It is calculated based on a set of assumed operating conditions well before the system approaches that operational state. The U.S. Federal Energy Regulatory Commission (FERC) requires the ATC to be periodically calculated and posted on the Open Access Same-time Information System (OASIS) for public access \cite{FERC96}.

In 1996, the North American Electric Reliability Council (NERC) proposed a uniform definition for ATC and the associated terminologies \cite{NERC96a}. 
By definition, ATC is a measure of the transfer capability remaining in the physical transmission network for further commercial activity over and above already committed uses. Mathematically, ATC is defined as:
\begin{equation}
\mbox{ATC = TTC -- TRM -- ETC -- CBM}
\label{eq:ATCdefinition}
\end{equation}
where TTC denotes the total transfer capability, TRM denotes the transmission reliability margin, ETC denotes the existing transmission commitments (base case) which include retail customer service, and CBM denotes the capacity benefit margin (CBM). 
More detailed explanations for the terminologies are presented below.
\begin{itemize}
\item TTC is the amount of electric power that can be transferred over the interconnected transmission network in a reliable manner while meeting all of a specific set of defined pre- and post-contingency system conditions (i.e., thermal, voltage, and stability limits). 
The TTC is the same as the first contingency total transfer capability (FCTTC) defined in \cite{NERC95a}.
\item TRM is the amount of transmission transfer capability necessary to ensure that the interconnected transmission network is secure under a reasonable range of \textbf{uncertainties} in system conditions. The TRM can be considered as the difference of TTC between cases with and without consideration of uncertainties brought about by, for instance, forecasted load demand and the integration of renewable energy resources (RES). 
TRM can be determined by assuming a fixed percentage reduction in TTC (typically too conservative).
Reference \cite{PWSauer98a} summarized four methods to determine the TRM, of which the probabilistic approaches is preferable considering the time-variant system conditions.
\item CBM is the amount of transmission transfer capability reserved by load-serving entities to ensure access to generation from interconnected systems to meet generation reliability requirements.
Typical CBM could be a multiple of the largest generation unit within the transmission system \cite{NERC99a}.
\item ETC is the sum of existing transfer capability between the source and the sink in the base case. 
\end{itemize}

Since we are interested in the probabilistic characteristics of the additional transfer capability on top of the base case (i.e., TTC -- ETC). Without loss of generality, we assume $\mbox{CBM}=0$ and use the term probabilistic ATC (PATC) to represent ATC plus TRM (i.e., TTC -- ETC) in the rest of the paper.

From the above definitions, it can be seen that the nucleus of assessing ATC is to calculate the PATC and its cumulative distribution function (CDF), based on which TRM can be estimated. As a result, the ATC can be computed directly from
\begin{equation}
\label{eq:ATCcompactdefinition}
\text{ATC} = \mathbb{E}[\text{PATC}] - \text{TRM}
\end{equation} 
Indeed, the main contribution of the paper is to propose a mathematical formulation for PATC and to develop a computationally efficient yet accurate algorithm for the PATC calculation.

In the previous work, traditional deterministic ATC calculation methods can be categorized into four classes: i) Linear approximation methods (e.g., \cite{GLLandgren72a}); ii) Repeated power flow (RPF) methods (e.g., \cite{YOu03a}); iii) Continuation power flow (CPF) methods (e.g., \cite{GCEjebe98a}, \cite{HDChiang05a}); and iv) Optimal power flow (OPF) methods (e.g., \cite{GDIrisarri97a}). %
However, deterministic ATC calculation does not consider the uncertainties of the renewable energy generation and their intrinsic dependencies.

To overcome the limitation of deterministic ATC methods, various probabilistic ATC assessment methods have been discussed in the previous literature, which can be categorized into 1) Monte-Carlo simulation; 2) Bootstrap method; 3) point estimation method; 4) stochastic programming.
Monte-Carlo simulation method is the most widely used method due to its simplicity. It has been used in combination with all deterministic ATC solvers mentioned above. However, it suffers from a high computational effort that may hamper its practical applications even with efficient sampling methods like the Latin Hypercube sampling method \cite{HYu09a}. 
Clustering methods \cite{MRamezani09a} were applied to speed up Monte-Carlo simulation yet at the expense of accuracy. 
The Bootstrap method is powerful in constructing confidence region of ATC but is time-consuming due to the re-sampling procedure \cite{RRChang02a}. 
Attempting to release the computational burden, some analytical methods have been developed utilizing mathematical approximations. In \cite{WYLi13a}, the point estimation method was proposed to estimate the variance of the TTC to determine the TRM and thus the ATC. 
Apart from the methods mentioned above, another class of methods for uncertainty quantification is the meta-modeling, which aims to build a statistically-equivalent functional representation for the desired response (e.g., the PATC in this study) using a small number of model evaluations. A representative method is the polynomial chaos expansion (PCE) which has been applied in the context of power systems to study the probabilistic power flow \cite{FNi17a}, the load margin problem \cite{EHaesen09a}, and the available delivery capability problem \cite{HSheng18a}. 
Another emerging method, an alternative to PCE, is the canonical low-rank approximations (LRA), which employs the canonical decomposition to express the desired response as a sum of rank-one functions \cite{KKonakli16a}. The original idea of canonical decomposition dates back to 1927 \cite{FHitchcock27a}, and the method has recently become attractive for uncertainty quantification in structural vibration problems \cite{MChevreuil15a}. 

In this paper, we propose a mathematical formulation of the PATC problem considering the uncertainties from the wind power, the solar PV and the loads. More importantly, we apply a novel LRA approach to solve the important PATC problem.
The main contribution of this paper are summarized as follows:
\begin{itemize}
\item A mathematical formulation of PATC problem is developed for transmission system integrating renewable energy resources and loads that may follow various marginal distributions. This formulation considers both the pre-contingency and post-contingency cases.
\item A novel efficient yet accurate algorithm---LRA is proposed to evaluate the PATC. 
Particularly, random variables with diverse marginal distributions and correlation can be accommodated using proper polynomial basis and Nataf transformation.
\item Accurate probabilistic characteristics (probabilistic density function (PDF), cumulative distribution function (CDF), mean, standard deviation, etc.) of the PATC can be achieved with much less computational efforts compared to the LHS-based MCS. Such probabilistic information can be further used to compute a more realistic TRM, leading to a more fully utilized transmission assets.  
\end{itemize}

The rest of the paper is organized as follows. Section II introduces the probabilistic models of the randomness in the PATC problem. Section III proposes the mathematical formulation of the PATC problem. Section IV elaborates the low-rank approximation method and its implementation in PATC calculation. The detailed algorithm to assess the PATC is summarized in Section V. The simulation results are presented in Section VI. Conclusions and perspectives are given in Section VII.

%
\section{Modeling Uncertainties in the Probabilistic ATC Problem}
\normalsize
The ATC calculation is typically applied to estimate the transfer capability for a pre-specified future period before the system approaches that operational state. To obtain a reasonable and dependable ATC, the base case condition, the target transactions, a credible contingency list, the physical and operational limits, and the network response should be well defined and estimated. 
The variability of diverse renewable energy sources and loads result in uncertainties in the base case, which therefore requires careful modeling. 

\subsection{Projection of the Wind Generation in Base Case}
\normalsize
In this study, the wind farms are modeled as an aggregated wind turbine with equivalent parameters. The Weibull distribution provides a great fit for the wind speed in many locations around the world \cite{SHJangamshetti99a}, the probability density function of which is 
\small
\begin{equation}
\label{eq:wind_pdf_weibull}
f_{V}(v)=\frac{k}{c}{{\left(\frac{v}{c}\right)}^{k-1}} \exp\left[-{{\left(\frac{v}{c}\right)}^{k}}\right]
\end{equation}
\normalsize
where ${v}$ is the wind speed, ${k}$ and ${c}$ are the equivalent shape and scale parameters, respectively.
As a result, the active power output ${{P}_{w}}$ can be calculated by the piece-wise wind speed-power output relation \cite{MAien15a} 
\small
\begin{equation}
\label{eq:wind_p_val}
{{P}_{w}}(v)=\left\{ \begin{array}{*{35}{l}}
0 & v\le {{v}_{in}} \mbox{ or } v>{{v}_{out}} \\
\displaystyle \frac{v-{{v}_{in}}}{{{v}_{rated}}-{{v}_{in}}}{{P}_{r}} & {{v}_{in}}<v\le {{v}_{rated}} \\
{{P}_{r}} & {{v}_{rated}}<v\le {{v}_{out}} \\
\end{array} \right.
\end{equation}
\normalsize
where ${{v}_{in}}$, ${{v}_{out}}$ and ${{v}_{rated}}$ are the cut-in, cut-out, and rated wind speed ($m/s$), ${{P}_{r}}$ is the rated wind power ($kW$). The reactive power can be determined according to the speed control type of the wind turbine \cite{EHCamm09b} since the wind turbine can be modelled as either a constant P-Q bus or a constant P-V bus with given Q-limits. 

\subsection{Projection of the Solar Generation in Base Case}
Similar to the wind farms, large solar PV power plants can be modeled as an aggregated solar PV with equivalent parameters. Typically, the solar radiation is represented by the Beta distribution \cite{ZMSalameh95a}
\small
\begin{equation}
\label{eq:solar_pdf_beta}
f_{R}(r)=\frac{\Gamma(\alpha+\beta)}{\Gamma(\alpha)\Gamma(\beta)}{{\left(\frac{r}{{{r}_{\max}}}\right)}^{\alpha-1}}{{\left(1-\frac{r}{{{r}_{\max}}}\right)}^{\beta-1}}
\end{equation}
\normalsize
where ${\alpha}$ and ${\beta}$ are the shape parameters of the distribution, ${\Gamma}$ denotes the Gamma function. The parameters are typically obtained from fitting historical solar radiation data. 
$r$ and ${{r}_{\max}}$ (${W/m}^{2}$) are the respective actual and maximum solar radiations. The active power ${{P}_{pv}}$ corresponding to the solar radiation ${r}$ is determined by the piece-wise function \cite{MAien15a} 
\small
\begin{equation}
\label{eq:solar_p_val}
{{P}_{pv}}(r)=\left\{ \begin{array}{*{35}{l}}
\displaystyle \frac{{{r}^{2}}}{{{r}_{c}}{{r}_{std}}}{{P}_{r}} & 0\le r<{{r}_{c}} \\
\displaystyle \frac{r}{{{r}_{std}}}{{P}_{r}} & {{r}_{c}}\le r \le {{r}_{std}} \\
{{P}_{r}} & r>{{r}_{std}} \\
\end{array} \right.
\end{equation}
\normalsize
where ${{r}_{c}}$ is a certain radiation point typically set as 150 $W/m^2$, ${{r}_{std}}$ is the solar radiation in the standard environment, ${{P}_{r}}$ is the rated power of the solar PV. Solar generation is required to be injected into the power grid at unity power factor \cite{WECC10}, and hence ${{Q}_{pv}}$ is assumed to be zero in this study.

\subsection{Projection of the Forecasted Load Demand in Base Case}

By nature, load demand is uncertain in power systems. 
It is a common practice to model the load uncertainty by Gaussian distribution due to the forecasting error \cite{RBillinton08a} 
\small
\begin{equation}
\label{eq:load_pdf}
f({{P}_{L}})=\frac{1}{\sqrt{2\pi}{{\sigma}_{P}}} \exp \left( -\frac{{{({{P}_{L}}-{{\mu}_{P}})}^{2}}}{2\sigma_{P}^{2}} \right)
\end{equation}
\normalsize
where the mean value ${{\mu}_{P}}$ of ${{P}_{L}}$ is provided by load forecaster, and ${{\sigma}_{P}}$ denotes the forecasting error. Generally, only the active power is predicted, whereas the reactive power is calculated under the assumption of constant power factor \cite{WYLi11a}.

%
\section{Mathematical Formulation of Probabilistic ATC Problem} \label{ATC_Formulation}

In this section, we present a CPF-based mathematical formulation of the probabilistic ATC problem. The near-future base case conditions are formulated as a probabilistic power flow problem in which uncertainties of the RES and loads are incorporated. Besides, target inter-area transactions are modeled as the load-generation variation vector.

The deterministic power flow equations of a $N$-bus transmission system can be represented as
\small
\begin{equation}
\label{eq:power_flow}
{\bm{f}}(\bm{x})=\begin{bmatrix}
{P_{Gi}-P_{Li}-P_{i}(\bm{x})} \\
{Q_{Gi}-Q_{Li}-Q_{i}(\bm{x})} \end{bmatrix}=0
\end{equation}
\normalsize
\noindent with
\small
\begin{equation}
\label{eq:bus_flow_out}
\begin{gathered}
P_{i}(\bm{x})=V_{i}\sum\limits_{j=1}^{N}{{V_{j}(G_{ij}\cos \theta_{ij}+B_{ij}\sin \theta_{ij})}} \\
Q_{i}(\bm{x})=V_{i}\sum\limits_{j=1}^{N}{{V_{j}(G_{ij}\sin \theta_{ij}-B_{ij}\cos \theta_{ij})}}
\end{gathered}
\end{equation}
\normalsize
where ${\bm{x}}={{[{\theta},{V}]}^{T}}$, e.g., voltage angles and magnitudes for all buses; ${P_{Gi}}$ and ${Q_{Gi}}$ are the total active and reactive generation power at bus $i$; ${P_{Li}}$ and ${Q_{Li}}$ are the total active and reactive load power at bus $i$; ${G_{ij}}$ and ${B_{ij}}$ are the real and imaginary part of the entry ${Y_{ij}}$ in the bus admittance matrix. 

Let ${\bm{v}}$, ${\bm{r}}$ and ${\bm{P}_{L}}$ be the random vectors representing wind speeds, solar radiations and load variations in the pre-specified period, respectively. The resulting base case condition of a ${N}$-bus system can be described as a set of probabilistic power flow (PPF) equations. Specifically, for P-Q type buses, the PPF equations are:
\small
\begin{equation}
\label{eq:pq_bus_pq}
\begin{gathered}
P_{Gi}+P_{wi}({{v}_{i}})+P_{pvi}({{r}_{i}})-P_{Li}(P_{Li})-P_{i}(\bm{x})=0 \\
Q_{Gi}+Q_{wi}({{v}_{i}})-Q_{Li}(P_{Li})-Q_{i}(\bm{x})=0
\end{gathered}
\end{equation}
\normalsize
\noindent For P-V type buses, the corresponding PPF equations are:
\small
\begin{equation}
\label{eq:pv_bus_pvq}
\begin{gathered}
P_{Gi}+P_{wi}({{v}_{i}})+P_{pvi}({{r}_{i}})-P_{Li}(P_{Li})-P_{i}(\bm{x})=0 \\
V_{i}={{V}_{i0}} \\
Q_{Gi}=-Q_{wi}({{v}_{i}})+Q_{Li}(P_{Li})+Q_{i}(\bm{x}) \\
{{Q}_{min,i}}\le Q_{Gi}\le {{Q}_{max,i}}
\end{gathered}
\end{equation}
\normalsize
\noindent where ${P_{wi}({{v}_{i}})}$, ${P_{pvi}({{r}_{i}})}$, ${P_{Li}}$ and ${P_{Gi}}$ are the real power injection from the wind farm, the solar PV power plant, the load, and the conventional generator at bus ${i}$; ${Q_{wi}({{v}_{i}})}$, ${Q_{Li}}$ and ${Q_{Gi}}$ are the corresponding reactive power injections.
If ${Q_{Gi}}$ exceeds its limits, then the terminal bus switches from P-V to P-Q with ${Q_{Gi}}$ fixed at the violated limit.

Besides, the target transactions under the study can be described as a load-generation variation vector ${\bm{b}}$ of the system in the form of
\small
\begin{equation}
\label{eq:cpf_variation}
{\bm{b}}=\begin{bmatrix}
\Delta{\bm{P}}_{G}-\Delta{\bm{P}}_{L} \\
-\Delta{\bm{Q}}_{L}
\end{bmatrix}
\end{equation}
\normalsize

In order to study how much power can be transferred along the direction specified by vector $b$, the set of equations (\ref{eq:pq_bus_pq})-(\ref{eq:cpf_variation}) can be formulated as a set of probabilistic CPF equations in the following compact form: 
\small
\begin{equation}
\label{eq:cpf_equation}
{\bm{f}}({\bm{x},\bm{\mu},\bm{\lambda},\bm{U}})={\bm{f}}({\bm{x},\bm{\mu}},{\bm{U}})-\lambda{\bm{b}}=0 
\end{equation}
\normalsize
\noindent where ${\bm{x}}$ is the vector of state variables, ${\bm{\mu}}$ is the vector of control parameters such as the tap ratio of transformers, ${\bm{U}=[\bm{v},\bm{r},\bm{{P}_{L}}]}$ is the random vector describing the wind speed, the solar radiation, and the load active power. It is obvious that the set of the parameterized power flow equations become the base-case power flow equation if ${\lambda=0}$.

For reliable operation of a power system, ATC calculation is required to account for both normal operating state and the state when a contingency occurs. Typically, only the ${N-1}$ contingencies are of interest in ATC calculation. However, more complex multiple contingencies may be required by the reliability criteria in a certain individual system. Enumeration of all contingencies is unnecessary and usually leads to a too conservative ATC value, hence it is a common practice to obtain a credible contingency list from the Security Analysis (SA) module in the Energy Management System (EMS). Therefore, the probabilistic ATC formulation considering a credible contingency list can be formulated as below:
\small
\begin{equation}
\label{eq:atc_final}
\lambda_{ATC} = min\{\lambda^{(0)},\lambda^{(1)},...,\lambda^{(N_{C})}\}
\end{equation}
\normalsize
\noindent in which the transfer capability $\lambda^{(v)}$ for the $v$-th case is determined by
\small
\begin{equation}
\label{eq:atc_equation}
\begin{aligned}
& \text{maximize} & & \lambda^{(v)} \\
& \text{subject to} & & {\bm{f}^{(v)}}({\bm{x}},{\bm{\mu}},{\bm{U}})-\lambda^{(v)}{\bm{b}}=0, & (a) \\
& & & {{V}_{min,i}^{(v)}}\le V_{i}({\bm{x},\bm{\mu},\lambda^{(v)},\bm{U}})\le {{V}_{max,i}^{(v)}}, & (b) \\
& & & S_{ij}^{(v)}({\bm{x},\bm{\mu},\lambda^{(v)},\bm{U}})\le {{S}_{ij,max}^{(v)}}, & (c) \\
& & & {{P}_{min,i}}\le P_{Gi}({\bm{x},\bm{\mu},\lambda^{(v)},\bm{U}})\le {{P}_{max,i}}, & (d) \\
& & & {{Q}_{min,i}}\le Q_{Gi}({\bm{x},\bm{\mu},\lambda^{(v)},\bm{U}})\le {{Q}_{max,i}}, & (e) \\
\end{aligned}
\end{equation}
\normalsize
where $v=0$ represents the base case, and $v=1,...,{{N}_{C}}$ represent the contingency cases. Specifically, ${\bm{f}^{(0)}}$ corresponds to the pre-contingency (base case) network configuration, while ${\bm{f}^{(v)}},v=1,...,{{N}_{C}}$ corresponds to the $v$-th post-contingency network configuration.
Similarly, $[{{V}_{min,i}^{(0)}},{{V}_{max,i}^{(0)}}]$ and ${{S}_{ij,max}^{(0)}}$ are the respective normal voltage and thermal limits applied to the base case; $[{{V}_{min,i}^{(v)}},{{V}_{max,i}^{(v)}}]$ and ${{S}_{ij,max}^{(v)}} (v=1,...,{{N}_{C}})$ are the emergency voltage and thermal limits applied to the contingency cases.
${\lambda}$ is the normalized load margin under the given load-generation variation vector.
The maximum value of $\lambda$ that could be achieved without the violation of (\ref{eq:atc_equation}) gives the ATC.
Note that ${\lambda}$ is a random variable due to the random input ${\bm{U}}$. Equation ($a$) specifies that the solution must satisfy the parameterized power flow equations (\ref{eq:cpf_equation}); Equations ($b$)-($e$) imply that the solution has to satisfy typical operational and electrical constraints.

For each realization of ${\bm{U}}$, there are $N_{C}+1$ ($N_{C}$ is the size of the contingency list) deterministic cases of (\ref{eq:atc_equation}) to be solved. As a result, it is extremely time-consuming to run Monte-Carlo simulations to estimate PATC using a large number of samples. It is imperative and essential to develop an efficient and accurate method to estimate the PATC.


%
\section{Canonical Low-rank Approximation using Polynomial Basis} \label{LRA_Section}
This section presents a general framework of the low-rank approximation of a multivariate stochastic response function. For simplicity, we first consider a scalar response function of independent inputs. The case of dependent inputs will be addressed in Section \ref{LRA_Section}--E.

\subsection{Low-rank Approximation with Polynomial Basis}

Consider a random vector ${\bm{\xi}}$ = (${{\xi}_{1},{\xi}_{2},...,{\xi}_{n},}$) with joint probability density function (PDF) ${f_{\bm{\xi}}}$ and marginal distribution functions ${f_{\xi_{i}},i=1,...,n}$ (${\xi_{i}}$ is related with the random variables ${U_{i}}$ in \eqref{eq:atc_equation}, see Section \ref{ATC_Formulation}), then the canonical rank-$r$ approximation \cite{KKonakli16a} of the target stochastic response (e.g., \textbf{PATC in this study}) ${Y = g(\bm{\xi})}$ can be represented by:
\small
\begin{equation}
\label{eq:lra_rank_R}
{Y}\approx{\hat{Y}}=\hat{g}(\bm{\xi})=\sum\limits_{l=1}^{r}{{{b}_{l}}{{\omega}_{l}}(\bm{\xi})}
\end{equation}
\normalsize
\noindent in which $b_{l},l=1,...,r$ are normalizing weighting factors, and ${{\omega}_{l}}$ is a rank-one function of ${\xi}$ in the form of 
\small
\begin{equation}
\label{eq:lra_rank_one}
\omega_{l}(\bm{\xi})=\prod\limits_{i=1}^{n}{{{v}_{l}^{(i)}}({{\xi}_{i}})}
\end{equation}
\normalsize
\noindent where ${v_{l}^{(i)}}$ denotes the ${i}$-th dimensional univariate function in the ${l}$-th rank-one function.
For most applications, the number ${r}$ of rank-one terms is usually small (under 5), hence \eqref{eq:lra_rank_R} and \eqref{eq:lra_rank_one} represent a canonical low-rank approximation.

In order to obtain the rank-$r$ approximation, a natural choice is expanding ${v_{l}^{(i)}}$ on a polynomial basis ${\{\phi_{k}^{(i)}, k \in N\}}$ that is orthogonal to ${f_{X_{i}}}$, the resulting rank-$r$ approximation takes the form:
\small
\begin{equation}
\label{eq:lra_rank_r_pce}
{\hat{Y}}=\hat{g}(\bm{\xi})=\sum\limits_{l=1}^{r}{{{b}_{l}} \left[\prod\limits_{i=1}^{n}{\left(\sum\limits_{k=0}^{{{p}_{i}}}{z_{k,l}^{(i)}\phi_{k}^{(i)}({{\xi}_{i}})} \right)} \right]}
\end{equation}
\normalsize
where ${\phi}_{k}^{(i)}$ denotes the $k$-th degree univariate polynomial in the $i$-th random input, $p_{i}$ is the maximum degree of ${\phi}^{(i)}$ and ${z}_{k,l}^{(i)}$ is the coefficient of ${\phi}_{k}^{(i)}$ in the $l$-th rank-one function.

Building the low-rank approximation for desired response in \eqref{eq:lra_rank_r_pce} requires: (i) choose an appropriate univariate polynomial for each random input; (ii) solve the polynomial coefficients ${z_{k,l}^{(i)}}$ as well as the weighing factors ${b_{l}}$. 
This process relies on a set of samples and their corresponding accurate response which are usually termed as experimental design (ED). 

\subsection{Selection of the Univariate Polynomial Basis}

It is crucial to choose a proper polynomial ${\phi}_{i}$ for the $i$th random input ${\xi}_{i},i=1,2,...,n$ to avoid low convergence rate and/or higher degree of expansion \cite{DBXiu02a}, which will hamper the capability of LRA in dealing with high-dimensional problems.
Table \ref{tab:gpce_mapping} shows a set of typical continuous distributions and the respective optimal Wiener-Askey polynomial basis, 
which can ensure the exponential convergence rate. In case that $f_{\xi_{i}}$ is out of the list in Table \ref{tab:gpce_mapping}, the discretized Stieltjes procedure is adopted to numerically construct a set of univariate orthogonal polynomial basis 
\cite{UQLabPCE17}. 
\begin{table}[t]
\renewcommand{\arraystretch}{1.3}
\caption{Standard Forms of Classical Continuous Distributions and their Corresponding Orthogonal Polynomials \cite{DBXiu02a}}
\label{tab:gpce_mapping}
\begin{center}
\begin{tabular}{c|c|>{\centering}p{1.5cm}|c}
\hline
\bfseries Distribution & \bfseries Density Function & \bfseries Polynomial & \bfseries Support \\
\hline
Normal & ${\frac{1}{\sqrt{2\pi}}{{e}^{-{{x}^{2}}/2}}}$ & Hermite & (-$\infty$,$\infty$) \\
\hline
Uniform & ${\frac{1}{2}}$ & Legendre & [-1,1] \\
\hline
Beta & ${\frac{{{(1-x)}^{\alpha}}{{(1+x)}^{\beta}}}{{{2}^{\alpha+\beta+1}}{B(\alpha+1,\beta+1)}}}$ & Jacobi & [-1,1] \\
\hline
Exponential & ${{e}^{-x}}$ & Laguerre & (0,$\infty$) \\
\hline
Gamma & ${\frac{{{x}^{\alpha}}{{e}^{-x}}}{\Gamma(\alpha+1)}}$ & Generalized Laguerre & [0,$\infty$) \\
\hline
\end{tabular}
\end{center}
\begin{tablenotes}
\item * The Beta function is defined as $B(p,q)=\frac{\Gamma(p)\Gamma(q)}{\Gamma(p+q)}$.
\end{tablenotes}
\end{table}
\subsection{Calculation of the Coefficients and Weighing Factors}\label{section_solve_lra}

Based on an experiment design of size $M_{C}$, i.e., a set of samples of $\bm{\xi}_{C}=\{\xi^{(1)},\xi^{(2)},...,\xi^{(M_{C})}\}$ and the corresponding response $\bm{y}_{C}=\{y^{(1)},y^{(2)},...,y^{(M_{C})}\}$ evaluated by deterministic tools, different algorithms have been proposed in the literature for solving the LRA coefficients and the weighting factors in a non-intrusive manner \cite{ADoostan13a}, \cite{PRai14a}. 
The sequential correction-updating scheme presented in \cite{MChevreuil15a} is employed in this study due to its efficiency and capability of constructing low-rank approximation using less sample evaluations. 
In the ${r}$-th correction step, a new rank-one function ${\omega_{r}}$ is built, while in the ${r}$-th updating step, the set of weighing factors ${\{b_{1},...,b_{r}\}}$ is determined. This process continues until the applied error index (the relative empirical error in this study) stop decreasing \cite{MChevreuil15a}.

\textbf{Correction step}:
the ${r}$-th correction step aims to find a new rank-one tensor $\omega_{r}$, which can be obtained by solving the following minimization problem:
\small
\begin{equation}
\label{eq:lra_rank_one_min}
\begin{aligned}
{{\omega}_{r}}(\bm{\xi}) &= \arg \underset{\omega \in W}{\mathop{\min}}\,\left\| {{e}_{r-1}}-\omega \right\|_{\bm{\xi}_{C}}^{2} \\
& = \arg \underset{\omega \in W}{\mathop{\min}}\,\sum_{m=1}^{M_{C}}{\left[y^{(m)}-\hat{g}_{r-1}(\bm{\xi}^{(m)})-\omega(\bm{\xi}^{(m)})\right]^{2}}
\end{aligned}
\end{equation}
\normalsize
\noindent where ${W}$ represents the space of rank-one tensors, ${{e}_{r-1}}=(g-{{\hat{g}}_{r-1}})$ is the approximation error of the response $Y$ at the ${(r-1)}$-th step, $\|.\|^{2}$ represents the norm 2 of the residual after the new rank-one tensor $w$ is applied, and the subscript ${\bm{\xi}_{C}}$ indicates that the minimization is carried over the whole set of samples in the experiment design $(\bm{\xi}_{C},\bm{y}_{C})$.

By exploiting the retained tensor-product form of the univariate polynomial basis, as shown in \eqref{eq:lra_rank_r_pce}, typical scheme for solving equation \eqref{eq:lra_rank_one_min} is the alternated least-square (ALS) minimization, which involves sequential minimization along each dimension ${i=1,...,n}$ to solve the corresponding polynomial coefficients ${\bm{z}_{r}^{(i)}=(z_{0,r}^{(i)},...,z_{p_{i},r}^{(i)})}$.
The total number of coefficients to be solved in each correction step is $\sum\nolimits_{i=1}^{n}{\left({{p}_{i}}+1\right)}$, which grows linearly as the number of random inputs $n$ increases.
Since ${{\omega }_{r}}$ is the product of $v_{l}^{(i)}(\xi)$ as shown in \eqref{eq:lra_rank_one}, $v_{l}^{(i)}({\xi}_{i})$ can be initialized by setting to $1.0$.

In the minimization along the ${i}$-th dimension, the polynomial coefficients corresponding to all other dimensions are "frozen" at their current values and the coefficients ${\bm{z}_{r}^{(i)}=(z_{0,r}^{(i)},...,z_{p_{i},r}^{(i)})}$ can be determined by:
\small
\begin{equation}
\label{eq:lra_coeff}
\bm{z}_{r}^{(i)}=\arg \underset{\bm{\zeta} \in {{R}^{({{p}_{i}}+1)}}}{\mathop{\min}}\,\left\|{{e}_{r-1}}-C_{i}\left( \sum\limits_{k=0}^{{{p}_{i}}}{{{\zeta}_{k}}\phi_{k}^{(i)}}\right)\right\|_{\bm{\xi}_{C}}^{2}
\end{equation}
\normalsize
\noindent where $C_{i}$ is a scalar
\small
\begin{equation}
\label{eq:lra_coeff_c}
C_{i}=\prod\limits_{j\ne i}{v_{r}^{(j)}(\xi_{j})}=\prod\limits_{j\ne i}{\left( \sum\limits_{k=0}^{{{p}_{j}}}{z_{k,r}^{\left( j \right)}\phi _{k}^{\left( j \right)}\left( {{\xi}_{j}} \right)} \right)}
\end{equation}
\normalsize

\textbf{Updating step}: After the ${r}$-th correction step is completed, the algorithm proceed to the ${r}$-th updating step to determine the weighing factor $b_{r}$ of the newly solved rank-one function ${{\omega }_{r}(\xi)}$, meanwhile, the set of existing weighing factors ${\bm{b}=(b_{1},...,b_{r-1})}$ are updated too. The updating step can be achieved by solving the following minimization problem
\small
\begin{equation}
\label{eq:lra_weight}
\bm{b}=\arg \underset{\bm{\beta} \in {{R}^{r}}}{\mathop{\min}}\,\left\|g-\sum\limits_{l=1}^{r}{{{\beta}_{l}}{{\omega}_{l}}}\right\|_{\bm{\xi}_{C}}^{2}
\end{equation}
\normalsize

\textbf{Stop criteria}: The correction-updating scheme successively adds new rank-one function to improve the accuracy of the approximation \eqref{eq:lra_rank_r_pce}. Hence, error reduction in two successive iterations becomes a natural stop criteria for this process. In this study, the relative empirical error is employed which is given by:
\small
\begin{equation}
\label{eq:lra_error}
{{\hat{e}}_{r}}=\frac{\left\|{{e}_{r-1}}-{{\omega}_{r}}\right\|_{\bm{\xi}_{C}}^{2}}{\mathbb{V}(\bm{y}_{C})}
\end{equation}
\normalsize
where $\mathbb{V}(\bm{y}_{C})$ denotes the empirical variance of the desired response over the experimental design.

Remark: (i) It is worth pointing out that the minimization problems in (\ref{eq:lra_coeff}) and (\ref{eq:lra_weight}) can be efficiently solved with the ordinary least-squares (OLS) technique because the dimension of unknowns are small.
(ii) When the LRA \eqref{eq:lra_rank_r_pce} for the desired response is built up, the response of any new samples can be evaluated efficiently by directly substituting to \eqref{eq:lra_rank_r_pce} instead of solving the original complex problem (e.g., the PATC problem \eqref{eq:atc_equation}).

\subsection{Selection of optimal rank and polynomial degree}
Currently, there is no systematic way to identify the optimal rank $r$ and the polynomial degree $p_i$ for individual random input beforehand. In this study, we specify a candidate set ${\{1,2,3,4,5\}}$ for the rank $r$, and another one ${\{2,3,4,5\}}$ for all the univariate polynomials. The rank selection is performed by progressively increasing the rank and applying the corrected error \eqref{eq:lra_error} to select the best one. The selection of optimal degree can be implemented similarly.

\subsection{Integration of Dependent Random Inputs}

So far, the random inputs are assumed to be mutually independent as requested by the LRA method.
To accommodate dependent random inputs with correlation matrix $\bm{\rho}$, the Nataf transformation \cite{RLebrun09a}, \cite{RLebrun09b} and the isoprobabilistic transformation can be employed to build up a mapping between $\bm{U}$ and the independent standard random variables $\bm{\xi}$: $\bm{u}=T(\bm{\xi})$, where $T$ is invertible. Therefore, the set of samples of ${\bm{\xi}}$ can be transformed back into samples of ${\bm{U}}$ to evaluate the corresponding responses ${\bm{Y}}$, after which the desired response $\bm{y}=g(T^{-1}(\bm{u}))$ can be expanded onto the polynomial basis with $\bm{\xi}$ using the aforementioned method \cite{KKonakli16a}.

\subsection{Moments of a Low-Rank Approximation}

Due to the orthogonality of the univariate polynomials that form the LRA basis (see (\ref{eq:lra_rank_r_pce})), the mean and the variance of the meta-model can be obtained \textit{analytically} in terms of the polynomial coefficients and the weighting factors. In particular, the mean and variance of the LRA response are given respectively by \cite{KKonakli16a}:
\small
\begin{equation}
\label{eq:lra_mean_var}
\begin{gathered}
{{\mu}_{y}}=E\left[{\hat{g}}(\bm{\xi})\right]=\sum\limits_{l=1}^{r}{{{b}_{l}}\left(\prod\limits_{i=1}^{n}{z_{0,l}^{(i)}}\right)} \\
\sigma_{y}^{2}=\sum\limits_{l=1}^{r}{\sum\limits_{m=1}^{r}{{{b}_{l}}{{b}_{m}}\prod\limits_{i=1}^{n}{\left[\left(\sum\limits_{k=0}^{{{p}_{i}}}{z_{k,l}^{(i)}z_{k,m}^{(i)}}\right)-z_{0,l}^{(i)}z_{0,m}^{(i)} \right]}}}
\end{gathered}
\end{equation}
\normalsize
Hence, if only mean and variance are of interests, \eqref{eq:lra_mean_var} can be applied directly without evaluating a large size of samples, which in contrast is required by most of the simulation-based methods.

%
\section{Computation of Probabilistic Available Transfer Capability}

In this section, a step-by-step description of the LRA method for PATC calculation is summarized below:

\noindent \textbf{Step 1}: Input the network data, the transactions to be studied, the credible contingency list, the probability distribution and the parameters of the random inputs $\bm{U}$, i.e., the wind speed, the solar radiation, the active load power, and their correlation matrix ${\bm{\rho}}$.

\noindent \textbf{Step 2}: Build the load-generation pattern (i.e., vector $b$ in \eqref{eq:atc_equation}-(a)) according to the transactions under study.

\noindent \textbf{Step 3}: Choose the independent standard variable ${\xi_{i}}$ and the corresponding univariate polynomial ${\phi_{i}}$ for each random input $U_{i}$.

\noindent \textbf{Step 4}: Generate an experimental design of size ${{M}_{C}}$: 
\begin{itemize}
\item i) Generate ${M_{C}}$ samples ${\bm{\xi}_{C}=(\xi^{(1)},\xi^{(2)},...,\xi^{({M_{C}})})}$ in the standard space by the LHS method.
\item ii) Transform ${\bm{\xi}_{C}}$ into the physical space by the inverse Nataf transformation ${\bm{u}_C}=T^{-1}(\bm{\xi}_C)$.
\item iii) Apply the deterministic ATC solver to evaluate the accurate response ${\bm{y}_{C}}$ (i.e., PATC) of ${\bm{u}_{C}}$. 
For each realization, the overall ATC is computed by applying the parallel scheme \cite{GCEjebe98a} which enables an early stop in CPF calculation when the first violation is encountered in either normal or contingency cases.
Pass the sample-response pairs $({\bm{\xi}_{C}},{\bm{y}_{C}})$ to Step 5.
\end{itemize}

\noindent \textbf{Step 5}: Apply the algorithm in Section \ref{section_solve_lra} to build the low-rank approximation (\ref{eq:lra_rank_r_pce}) for the PATC.
If the LRA for PATC has reached the prescribed accuracy, go to Step 7; otherwise, go to Step 6.

\noindent \textbf{Step 6}: Generate additional ${\Delta M_{C}}$ new samples and evaluate them, then go back to Step 5 using the enriched experiment design $({\bm{\xi}_{C}}+{\Delta \bm{\xi}_{C}},{\bm{y}_{C}}+{\Delta \bm{y}_{C}})$.

\noindent \textbf{Step 7}: Calculate the mean and standard deviation of the response through \eqref{eq:lra_mean_var}. 

\noindent \textbf{Step 8}: Sample $\bm{\xi}$ extensively, e.g., ${{M}_{S}}$ samples, and apply the solved functional approximation (\ref{eq:lra_rank_r_pce}) to evaluate the corresponding response $\bm{y}_{S}$ for all samples. Then compute the statistics of interest (PDF/CDF of PATC in this study).

\noindent \textbf{Step 9}: Compute the TRM value and the resulting ATC ($ATC=\mu_{PATC} - TRM$) for the given confidence level $p_{cl}\%$.

\noindent \textbf{Step 10}. Generate the result report.

Remark: the number of samples ${{M}_{C}}$ in Step 4 is usually much smaller than ${{M}_{S}}$ in Step 8. Unlike MCS, LRA does not solve power flow equations for all ${{M}_{S}}$ samples in Step 8, and hence it is more efficient. The main computational effort of LRA lies in Step 4.

%
\section{Numerical Studies}

In this section, we apply the proposed LRA method to investigate the probabilistic ATC of the modified IEEE 24-bus reliability test system (RTS) 
and IEEE 118-bus systems \cite{MATPOWER}. The LHS-based Monte Carlos simulation serves as a benchmark for validating the accuracy and the performance of the proposed method.
In addition, a comparison between the LRA and the sparse PCE \cite{HSheng18a} is also presented.

In this study, we assume that the probability distributions and the associated parameters of all random inputs are available from up-front modeling. Particularly, the wind speed follows Weibull distribution; the solar radiation follows Beta distribution; the load power follows Normal distribution. For each individual load, the mean value is set to be its base case value and the variance is equal to 5\% of its mean value. 
The univariate polynomial basis used in \eqref{eq:lra_rank_r_pce} for Weibull, Beta and Normal distributions are chosen appropriately 
according to Table \ref{tab:gpce_mapping}.
For simplicity, the linear correlation coefficient ${\rho_{ij}}$ between component $i$ and $j$ of wind speed $\bm{v}$, solar radiations $\bm{r}$ and load power $\bm{P_{L}}$ are 0.8040, 0.5053 and 0.4000, respectively.

\subsection{The Modified IEEE 24-Bus RTS}

The IEEE 24-bus RTS is composed of 4 areas, containing 33 generators, 32 branches and 17 loads. 
In this paper, we modify this test system by adding 4 wind farms of 80 MW at bus \{15, 18, 21, 23\} and 4 solar PV power plants of 60 MW at bus \{1, 2, 7, 16\}, respectively. Their corresponding parameters $\{W1, W2, W3, W4\}$ and $\{S1, S2, S3, S4\}$ are shown in Table \ref{tab:wind_param}-\ref{tab:solar_param}. Together with 17 stochastic loads, there are totally 25 random inputs. 
The existing transmission commitments (base case) is the first-day peak load as defined in \cite{PFAlbrecht79a}, and the new transaction under study is to transfer 75 MW from generator bus \{7\} in area 2 (source) to load buses \{3,4,9\} in area 1 (sink). The contingency list under study contains four $N-1$ outages $\{G1\#1, L2-4, L3-24, L9-11\}$.

\begin{table}[]
\renewcommand{\arraystretch}{1.3}
\caption{Wind speed and wind turbine parameters \cite{SHJangamshetti99a}}
\label{tab:wind_param}
\centering
\begin{tabular}{c|c|c|c|c|c}
\hline
$No.$ & ${c}$ & ${k}$ & ${V_{r}}$ & ${V_{in}}$ & ${V_{out}}$ \\
\hline
${W1}$ &  8.0063 & 2.1182 & 13.50 & 3.50 & 25.00 \\
\hline
${W2}$ & 11.5762 & 2.7022 & 13.80 & 3.50 & 25.00 \\
\hline
${W3}$ & 11.2441 & 3.6322 & 13.00 & 5.00 & 25.00 \\
\hline
${W4}$ & 12.4813 & 3.2465 & 12.90 & 5.00 & 24.00 \\
\hline
${W5}$ & 11.1533 & 3.2895 & 12.00 & 5.50 & 24.00 \\
\hline
${W6}$ &  8.8261 & 2.6511 & 10.00 & 3.50 & 20.00 \\
\hline
\end{tabular}
\end{table}
\begin{table}[]
\renewcommand{\arraystretch}{1.3}
\caption{Solar radiation and solar PV parameters \cite{FYEttoumi02a}}
\label{tab:solar_param}
\centering
\begin{tabular}{c|c|c|c|c|c|c}
\hline
$No.$ & ${\alpha}$ & ${\beta}$ & $r_{min}$ & $r_{max}$ & ${R_{c}}$ & ${R_{std}}$ \\
\hline
${S1}$ & 1.110 & 0.730 & 0.0 & 1000.0 & 150.0 & 1000.0 \\
\hline
${S2}$ & 1.320 & 0.690 & 0.0 & 1000.0 & 150.0 & 1000.0 \\
\hline
${S3}$ & 1.700 & 0.740 & 0.0 & 1000.0 & 150.0 & 1000.0 \\
\hline
${S4}$ & 2.970 & 0.940 & 0.0 & 1000.0 & 150.0 & 1000.0 \\
\hline
${S5}$ & 2.540 & 0.780 & 0.0 & 1000.0 & 150.0 & 1000.0 \\
\hline
${S6}$ & 1.780 & 0.850 & 0.0 & 1000.0 & 150.0 & 1000.0 \\
\hline
\end{tabular}
\end{table}

We first apply the CPF-based ATC solver to the deterministic system (i.e., without uncertainty). As shown in Table \ref{tab:deter_atc}, the binding limit is the thermal limit at branch 7-8 in the base case, leading to an overall ATC 83.0207 MW. The contingencies do not deteriorate the transfer capability potentially because the given limits of branches are larger in emergency compared to the normal operating condition. 
\begin{table}[]
\renewcommand{\arraystretch}{1.3}
\caption{The determination of ATC considering normal and contingency cases}
\label{tab:deter_atc}
\centering
\begin{tabular}{c|c|>{\centering\arraybackslash}m{1.2cm}|>{\centering\arraybackslash}m{1.2cm}|>{\centering\arraybackslash}m{1.2cm}|>{\centering\arraybackslash}m{0.8cm}}
\hline
\multirow{2}{*}{Case} & \multirow{2}{*}{Outage} & \multicolumn{3}{c|}{ATC w.r.t limits (MW)} & \multirow{2}{*}{Overall} \\
\cline{3-5}
No. & Facility & Voltage limit & Thermal limit & Voltage collapse & ATC\\
\hline
${0}$ & Base case & 280.3934 & 83.0207 & 511.9046 & \multirow{5}{*}{83.0207} \\
\cline{1-5}
${1}$ & G1-1 & 469.0375 & 128.0214 & 508.8632 & \\
\cline{1-5}
${2}$ & L2-4 & 201.7742 & 127.8677 & 451.6651 & \\
\cline{1-5}
${3}$ & L3-24 & 183.7445 & 127.5457 & 355.2315 & \\
\cline{1-5}
${4}$ & L9-11 & 395.0070 & 127.8440 & 460.6714 & \\
\hline
\end{tabular}
\end{table}

\begin{figure}[h]
\centering
\includegraphics[width=0.45\textwidth,height=0.18\textheight]{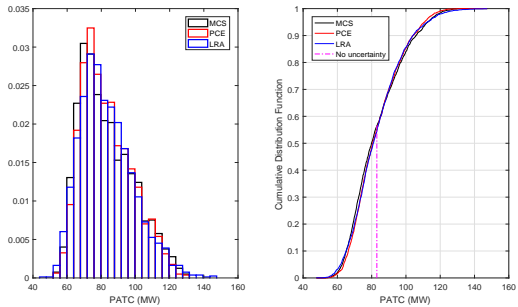}
\caption{The distribution of the PATC computed by the MCS, the PCE and the LRA. They are almost overlapped. The TRM for 95\% confidence level is 21.1497 MW and the resulting ATC is 61.8815 MW.}
\label{fig:total_atc_dist}
\end{figure}

Next, we exploit the proposed LRA method to assess the probabilistic characteristics (e.g., mean, variance, PDF and CDF) of the PATC and compare the results with those of the LHS-based MCS and with those of the PCE method.
Applying the proposed algorithm, 125 simulations are required in Step 4-5 to build up the LRAs \eqref{eq:lra_rank_r_pce} of the PATC, which consist of 1 rank-one function with the highest polynomial degrees $p_{i}=2$. The total number of coefficients plus weighting factors is 76 (i.e., $\sum\nolimits_{l=1}^{r}{\left(\sum\nolimits_{i=1}^{n}(p_{i}+1)+1\right)}=\sum\nolimits_{i=1}^{25}(2+1)+1$). Once the coefficients and the weighting factors are computed, the mean and the standard deviation of the responses are computed in Step 7 and are compared with those of the LHS-based MCS and with those of the sparse PCE method as shown in Table \ref{tab:comp_atc_24}. Furthermore, 2000 samples are generated in Step 8 to assess the PDF and the CDF of the response. Fig. \ref{fig:total_atc_dist} shows the PDF and CDF of the PATC computed by the LHS-based MCS, the sparse PCE, and the solved LRA, respectively. These results clearly demonstrate that the LRA-PATC method can provide accurate estimation for the probabilistic characteristics of the PATC. 
\begin{table}[]
\renewcommand{\arraystretch}{1.3}
\caption{Comparison of the estimated statistics of the overall ATC by the MCS, PCE and LRA methods}
\label{tab:comp_atc_24}
\centering
\begin{tabular}{c|c|c|c|c|c}
\hline
Indices & MCS & PCE & LRA & ${\frac{\Delta PCE}{MCS}} \%$ & ${\frac{\Delta LRA}{MCS}} \%$ \\
\hline
$\mu$ & 83.0312 & 83.2369 & 83.2226 & 0.2478 & 0.2305 \\
\hline
$\sigma$ & 15.5418 & 14.4898 & 15.4278 & -6.7692 & -0.7340 \\
\hline
\end{tabular}
\end{table}

Once we have the statistics of the PATC, a reasonable amount of TRM can be obtained from the CDF of PATC. Table \ref{tab:confid_levels_24} shows the TRM with different confidence levels and the resulting ATC values. For example, if 95\% confidence level is requested, i.e., $P(ATC_{actual} >= (\mu_{ATC} - TRM)) = 0.95$, the corresponding ATC is 61.8815 MW. 
\begin{table}[]
\renewcommand{\arraystretch}{1.3}
\caption{The estimated TRM and resulting ATC for different confidence levels}
\label{tab:confid_levels_24}
\centering
\begin{tabular}{>{\centering\arraybackslash}m{1.6cm}|>{\centering\arraybackslash}m{1.5cm}|>{\centering\arraybackslash}m{1.5cm}|>{\centering\arraybackslash}m{1.5cm}}
\hline
Confid. Level & $\mathbb{E}($PATC$)$ & TRM (MW) & ATC (MW) \\
\hline
${99.0\%}$ & 83.0312 & 26.2326 & 56.7986 \\
\hline
${98.0\%}$ & 83.0312 & 24.4386 & 58.5926 \\
\hline
${95.0\%}$ & 83.0312 & 21.1497 & 61.8815 \\
\hline
${90.0\%}$ & 83.0312 & 17.9604 & 65.0707 \\
\hline
${80.0\%}$ & 83.0312 & 12.9772 & 70.0540 \\
\hline
\end{tabular}
\end{table}
\subsection{The Modified 118-bus System}
The IEEE-118 bus system is a simplified representation of the Midwest U.S. transmission system in 1962, which contains 19 generators, 35 synchronous condensers, 177 transmission lines, 9 transformers and 91 loads. 
Six wind farms, each with $50$ MW, are connected to bus $\{10, 25, 26, 49, 65, 66\}$ using the parameters $\{W1, W2, W3, W4, W5, W6\}$ in Table \ref{tab:wind_param} respectively. Six solar PV parks, each with an installed capacity of $30$ MW, are connected to bus $\{12, 59, 61, 80, 89, 100\}$ using parameters $\{S1, S2, S3, S4, S5, S6\}$ in Table \ref{tab:solar_param}. Besides, there are 99 stochastic loads, leading to 111 random inputs in total. 
The total load in the base case is 4242 MW and 1438 Mvar. The new transaction under study is to transfer 60 MW from bus 89 to bus 91. The contingency list contains six $N-1$ outages $\{L88-89, L89-90, L90-91, L89-92, L91-92, L92-94\}$ surrounding the source and sink buses. 

Likewise, we first apply the CPF-based ATC solver to compute the maximum transfer without considering the uncertainties of RES and loads. The deterministic ATC is 24.1835 MW. Next, we exploit the proposed LRA-PATC method to assess the probabilistic characteristics of the PATC and compare the results with those of the LHS-based MCS and the sparse PCE method.

Applying the proposed algorithm, 556 simulations are needed in Step 4-5 to build up the LRAs \eqref{eq:lra_rank_r_pce} of the PATC, which consist of 1 rank-one function with the highest degree $p_{i}=2$. The total number of coefficients plus weighting factors is 334.
Eventually, the mean and standard deviation of the responses are computed in Step 7 and are compared with those from the LHS-based MCS and from the sparse PCE method as shown in Table \ref{tab:comp_atc_118}. 
All these results and comparisons clearly demonstrate that the LRA method can provide accurate estimation for the probabilistic characteristics of the PATC solutions.
However, to get comparable accuracy, the LHS-based MCS needs to run 10000 simulations (i.e., solving (\ref{eq:atc_equation})), taking 9110s, while the LRA takes only 480s, i.e., about $\frac{1}{19}$ of the computational time required by MCS. Clearly, the LRA is much more efficient than the MCS.
\begin{table}[]
\renewcommand{\arraystretch}{1.3}
\caption{Comparison of the estimated statistics of the overall ATC by the MCS, PCE and LRA methods}
\label{tab:comp_atc_118}
\centering
\begin{tabular}{c|c|c|c|c|c}
\hline
Indices & MCS & PCE & LRA & ${\frac{\Delta PCE}{MCS}} \%$ & ${\frac{\Delta LRA}{MCS}} \%$ \\
\hline
$\mu$ & 23.9278 & 23.7476 & 23.7550 & -0.7530 & -0.7220 \\
\hline
$\sigma$ & 4.8779 & 4.5220 & 4.8617 & -7.2973 & -0.3327 \\
\hline
\end{tabular}
\end{table}

The TRM and the resulting overall ATC for different confidence levels are shown in Table \ref{tab:confid_levels_118}, which empowers the system operator to determine a reasonable amount of TRM to make full use of the transmission assets. 
\begin{table}[]
\renewcommand{\arraystretch}{1.3}
\caption{The estimated TRM and resulting ATC for different confidence levels}
\label{tab:confid_levels_118}
\centering
\begin{tabular}{>{\centering\arraybackslash}m{1.6cm}|>{\centering\arraybackslash}m{1.5cm}|>{\centering\arraybackslash}m{1.5cm}|>{\centering\arraybackslash}m{1.5cm}}
\hline
Confid. Level & $\mathbb{E}($PATC$)$ & TRM (MW) & ATC (MW) \\
\hline
${99.0\%}$ & 23.9278 & 10.1839 & 13.7439 \\
\hline
${98.0\%}$ & 23.9278 & 8.9942 & 14.9336 \\
\hline
${95.0\%}$ & 23.9278 & 7.4975 & 16.4303 \\
\hline
${90.0\%}$ & 23.9278 & 6.1159 & 17.8119 \\
\hline
${80.0\%}$ & 23.9278 & 4.3121 & 19.6157 \\
\hline
\end{tabular}
\end{table}
%

%
\section{Conclusion and Perspectives}

In this paper, we have proposed a mathematical formulation for the probabilistic ATC (PATC) problem in which the uncertainties from RES and loads are incorporated. Moreover, we have proposed a novel LRA method which can assess the PATC accurately and efficiently by building up a statistically-equivalent low-rank approximation. 
Numerical studies show that the proposed method can accurately estimate the probabilistic characteristics of the PATC with much less computational effort compared to the LHS-based MCS. 

The PATC provides important insights into how the uncertainties may affect the transfer capability of the power system. More importantly, the proposed method can help determine the TRM and thus the ATC in a more efficient manner to fully utilize the transmission assets while maintaining the security and reliability of the grid.

In the future, we plan to develop control measures to reduce the variance of PATC to decrease the TRM for a full utilization of transmission assets.


\begin{thebibliography} {1}
%
\bibitem{FERC96}
Federal Energy Regulatory Commission, \textit{Open access same-time information system (formerly Real-Time Information Networks) and standards of conduct}. Docket No. RM 95-9-000, Order 889, 1996.
%
\bibitem{NERC96a}
Transmission Transfer Capability Task Force, \textit{Available Transfer Capability Definitions and Determination}. Princeton, NJ, USA: North American Electric Reliability Council, 1996.
%
\bibitem{NERC95a}
Transmission Transfer Capability Task Force, \textit{Transmission Transfer Capability}. Princeton, NJ, USA: North American Electric Reliability Council, 1995.
%
\bibitem{PWSauer98a}
P. W. Sauer, "Alternatives for calculating transmission reliability margin (TRM) in available transfer capability (ATC)," in \textit{31th Annual Hawaii International Conference on System Sciences}, Kohala Coast, HI, USA, 9--9 January, 1998.
%
\bibitem{NERC99a}
Available Transfer Capability Working Group, \textit{Transmission capability margins and their use in ATC determination-white paper}. Princeton, NJ, USA: North American Electric Reliability Council, 1999.
%
\bibitem{GLLandgren72a}
G. L. Landgren, H. L. Terhune, and R. K. Angel, "Transmission interchange capability-analysis by computer," \textit{IEEE Trans. Power App. Syst.}, vol. PAS-91, no. 6, pp. 2405--2414, 1972.
%
\bibitem{YOu03a}
Y. Ou and C. Singh, "Calculation of risk and statistical indices associated with available transfer capability," \textit{IEE Proc. Gener. Transm. Distrib.}, vol. 150, no. 2, pp. 239--244, 2003.
%
\bibitem{GCEjebe98a}
G. C. Ejebe, J. Tong, J. G. Waight, J. G. Frame, X. Wang, and W. F. Tinney, "Available transfer capability calculations," \textit{IEEE Trans. Power Syst.}, vol. 13, no. 4, pp. 1521--1527, 1998.
%
\bibitem{HDChiang05a}
H. Chiang and H. Li, "On-line ATC evaluation for large-scale power systems: framework and tool," in \textit{Applied Mathematics for Restructured Electric Power Systems}. New York, NY, USA: Springer, 2005, ch. 5, pp. 63--103.
%
\bibitem{GDIrisarri97a}
G. D. Irisarri, X. Wang, J. Tong, and S. Mokhtari, "Maximum loadability of power systems using interior point non-linear optimization method," \textit{IEEE Trans. Power Syst.}, vol. 12, no. 1, pp. 162--172, 1997.
%
\bibitem{HYu09a}
H. Yu, C. Y. Chung, K. P. Wong, and J. H. Zhang, "Probabilistic load flow evaluation with hybrid Latin Hypercube sampling and Cholesky decomposition," \textit{IEEE Trans. on Power Syst.}, vol. 24, no. 2, pp. 661--667, 2009.
%
\bibitem{MRamezani09a}
M. Ramezani, C. Singh, and M. Haghifam, "Role of clustering in the probabilistic evaluation of TTC in power systems including wind power generation," \textit{IEEE Trans. Power Syst.}, vol. 24, no. 2, pp. 849--858, 2009.
%
\bibitem{RRChang02a}
R. R. Chang, C. Y. Tsai, C. L. Su, and C. N. Lu, "Method for computing probability distributions of available transfer capability," \textit{IEEE Proc. Gener. Transm. Distrib.}, vol. 149, no. 4, pp. 427--431, 2002.
%
\bibitem{WYLi13a}
W. Li, E. Vaahedi, and Z. Lin, "BC Hydro's transmission reliability margin assessment in total transfer capability calculations," \textit{IEEE Trans. Power Syst.}, vol. 28, no. 4, pp. 4796--4802, 2013.
%
\bibitem{FNi17a}
F. Ni, P. H. Nguyen, and J. F. G. Cobben, "Basis-adaptive sparse polynomial chaos expansion for probabilistic power flow," \textit{IEEE Trans. Power Syst.}, vol. 32, no. 1, pp. 694--705, 2017.
%
\bibitem{EHaesen09a}
E. Haesen, C. Bastiaensen, J. Driesen, and R. Belmans, "A probabilistic formulation of load margins in power systems with stochastic generation," \textit{IEEE Trans. Power Syst.}, vol. 24, no. 2, pp. 951--958, 2009.
%
\bibitem{HSheng18a}
H. Sheng and X. Wang, "Applying polynomial chaos expansion to assess probabilistic available delivery capability for distribution networks with renewables," \textit{IEEE Trans. Power Syst.}, 2018, (Early Access).
%
\bibitem{KKonakli16a}
K. Konakli and B. Sudret, “Polynomial meta-models with canonical low-rank approximations: numerical insights and comparison to sparse polynomial chaos expansions,” \textit{J. Comput. Phys.}, vol. 321, pp. 1144--1169, 2016.
%
\bibitem{FHitchcock27a}
F. Hitchcock, "The expression of a tensor or a polyadic as a sum of products," \textit{J. Math. Phys.}, vol. 6, pp. 164--189, 1927.
%
\bibitem{MChevreuil15a}
M. Chevreuil, R. Lebrun, A. Nouy, and P. Rai, "A least-squares method for sparse low-rank approximation of multivariate functions," \textit{SIAM/ASA J.	Uncertain. Quantificat.}, vol. 3, no. 1, pp. 897--921, 2015.
%
\bibitem{SHJangamshetti99a}
S. H. Jangamshetti and V. G. Rau, "Site matching of wind turbine generators: a case study," \textit{IEEE Trans. Energy Conver.}, vol. 14, no. 4, pp. 1537--1543, 1999.
%
\bibitem{MAien15a}
M. Aien, M. Rashidinejad, and M. F. Firuz-Abad, "Probabilistic optimal power flow in correlated hybrid wind-PV power systems: a review and a new approach," \textit{Renewable \& Sustainable Energy Reviews}, vol. 41, pp. 1437--1446, 2015.
%
\bibitem{EHCamm09b}
E. H. Camm, M. R. Behnke, O. Bolado, M. Bollen, M. Bradt, C. Brooks, W.	Dilling, M. Edds, W. J. Hejdak, D. Houseman, S. Klein, F. Li, J. Li, P. Maibach, T. Nicolai, J. Patino, S. V. Pasupulati, N. Samaan, S. Saylors, T. Siebert, T. Smith, M. Starke, , and R. Walling, "Characteristics of wind turbine generators for wind power plants," in \textit{IEEE Power \& Energy Society General Meeting}, Calgary, AB, Canada, 26--30 July 2009, pp. 1--5.

%
\bibitem{ZMSalameh95a}
Z. M. Salameh, B. S. Borowy, and A. R. A. Amin, "Photovoltaic module-site matching based on the capacity factors," \textit{IEEE Trans. Energy Convers.}, vol. 10, no. 2, pp. 326--332, 1995.
%
\bibitem{WECC10}
WECC. WECC guide for representation of photovoltaic systems in large-scale load flow simulations. [Online]. Available: https://www.wecc.biz
%
\bibitem{RBillinton08a}
R. Billinton and D. Huang, "Effects of load forecast uncertainty on bulk electric system reliability evaluation," \textit{IEEE Trans. Power Syst.}, vol. 23, no. 2, pp. 418--425, 2008.
%
\bibitem{WYLi11a}
W. Li, \textit{Probabilistic transmission system planning}. Hoboken, NJ, USA: John Wiley \& Sons, Inc., 2011.
%
\bibitem{DBXiu02a}
D. B. Xiu and G. E. Karniadakis, "The Wiener-Askey polynomial chaos for stochastic differential equations," \textit{SIAM J. Sci. Comput.}, vol. 24, no. 2, pp. 619--644, 2002.
%
\bibitem{UQLabPCE17}
S. Marelli and B. Sudret, "UQLab user manual-polynomial chaos expansions," Chair of Risk, Safety \& Uncertainty Quantification, ETH Zurich, Tech. Rep. UQLab-V1.0-104, 2017.
%
\bibitem{ADoostan13a}
A. Doostan, A. Validi, and G. Iaccarino, "Non-intrusive low-rank separated approximation of high-dimensional stochastic models," \textit{Comput. Methods Appl. Mech. Eng.}, vol. 263, pp. 42--55, 2013.
%
\bibitem{PRai14a}
P. Rai, "Sparse low-rank approximation of multivariate functions-applications in uncertainty quantification," Ph.D. dissertation, Engineering Sciences [Physics], Ecole Centrale Nantes, 2014.
%
\bibitem{RLebrun09a}
R. Lebrun and A. Dutfoy, "An innovating analysis of the Nataf transformation from the copula viewpoint," \textit{Prob. Eng. Mech.}, vol. 24, no. 3, pp. 312--320, 2009.
%
\bibitem{RLebrun09b}
R. Lebrun and A. Dutfoy, "A generalization of the Nataf transformation to distributions with elliptical copula," \textit{Prob. Eng. Mech.}, vol. 24, no. 2, pp. 172--178, 2009.
%
\bibitem{MATPOWER}
R. D. Zimmerman, C. E. Murillo-Sanchez, and R. J. Thomas, "MATPOWER: steady-state operations, planning and analysis tools for power systems research and education," \textit{IEEE Trans. Power Syst.}, vol. 26, no. 1, pp. 12--19, 2011.
%
\bibitem{PFAlbrecht79a}
P. F. Albrecht, M. P. Bhavaraju, B. E. Biggerstaff, R. Billinton, G. E. Jorgensen, N. D. Reppen, and P. B. Shortley, "IEEE reliability test system," \textit{IEEE Trans. Power App. Syst.}, vol. 98, no. 6, pp. 2047--2054, 1979.
%
\bibitem{FYEttoumi02a}
F. Y. Ettoumi, A. Mefti, A. Adane, and M. Y. Bouroubi, "Statistical analysis of solar measurements in Algeria using beta distributions," \textit{Renewable Energy}, vol. 26, no. 1, pp. 47--67, 2002.
%
\end{thebibliography}
\end{document}